\pdfoutput=1
\documentclass[sigchi]{acmart}

\usepackage{verbatim}
\usepackage{graphicx}
\usepackage{subfig}
\usepackage{tikz}
\usepackage{microtype}
\usetikzlibrary{arrows,automata}

\def\BibTeX{{\rm B\kern-.05em{\sc i\kern-.025em b}\kern-.08emT\kern-.1667em\lower.7ex\hbox{E}\kern-.125emX}}
    
\copyrightyear{2019}
\acmYear{2019}
\setcopyright{iw3c2w3}
\acmConference[WWW '19 Companion]{Companion Proceedings of the 2019 World Wide Web Conference}{May 13--17, 2019}{San Francisco, CA, USA}
\acmBooktitle{Companion Proceedings of the 2019 World Wide Web Conference (WWW '19 Companion), May 13--17, 2019, San Francisco, CA, USA}
\acmPrice{}
\acmDOI{10.1145/3308560.3316754}
\acmISBN{978-1-4503-6675-5/19/05}

\begin{document}

%
\title[Understanding Wikipedia Articles through Motifs]{Understanding the Signature of Controversial Wikipedia Articles through Motifs in Editor Revision Networks}

%

\author{James R. Ashford}
\email{AshfordJR@cardiff.ac.uk}
\affiliation{
  School of Computer Science and Informatics, Cardiff University, \\UK
}

\author{Liam D. Turner}
\email{TurnerL9@cardiff.ac.uk}
\affiliation{
  School of Computer Science and Informatics, Cardiff University, \\UK
}

\author{Roger M. Whitaker}
\email{WhitakerRM@cardiff.ac.uk}
\affiliation{
  School of Computer Science and Informatics, Cardiff University, \\UK
}

\author{Alun Preece}
\email{PreeceAD@cardiff.ac.uk}
\affiliation{
  Crime and Security Research Institute, Cardiff University, \\UK
}

\author{Diane Felmlee}
\email{Dhf12@psu.edu}
\affiliation{
  Dept. of Sociology \& Criminology, Pennsylvania State University, \\USA
}

\author{Don Towsley}
\email{Towsley@cs.umass.edu}
\affiliation{
  College of Information and Computer Sciences, University of Massachusetts Amherst, USA
}

%
\renewcommand{\shortauthors}{Ashford et al.}

%
\begin{abstract}
Wikipedia serves as a good example of how editors collaborate to form and maintain an article. The relationship between editors, derived from their sequence of editing activity, results in a directed network structure called the revision network, that potentially holds valuable insights into editing activity. In this paper we create revision networks to assess differences between controversial and non-controversial articles, as labelled by Wikipedia. Originating from complex networks, we apply motif analysis, which determines the under or over-representation of induced sub-structures, in this case triads of editors. We analyse 21,631 Wikipedia articles in this way, and use principal component analysis to consider the relationship between their motif subgraph ratio profiles. Results show that a small number of induced triads play an important role in characterising relationships between editors, with controversial articles having a tendency to cluster. This provides useful insight into editing behaviour and interaction capturing counter-narratives, without recourse to semantic analysis. It also provides a potentially useful feature for future prediction of controversial Wikipedia articles. 
\end{abstract}


%

\maketitle
\section{Introduction}

Wikipedia has become a tremendous platform for crowdsourcing knowledge, representing a cornerstone of the World Wide Web \cite{crowd}. It allows the "wisdom of the crowd'' \cite{Surowiecki2005} to potentially emerge, providing intelligence on a vast range of topics \cite{crowd_intel}. However, complex dynamics support the emergence of content, since the formation of Wikipedia articles involves both human cooperation and human conflict, based on the extent of convergent and divergent views. Narrative and counter-narrative frequently jostle for presence in a articles, representing a source of friction that is seen through editor interaction \cite{wiki2} and in the semantics of article content \cite{wiki3}. Wikipedia conveniently provides a list of controversial content that are labelled by the Wikipedia community themselves\footnote{\url{https://en.wikipedia.org/wiki/Wikipedia:List_of_controversial_issues}}. 

In an age of misinformation \cite{mis1, mis2, mis3}, understanding characteristics of controversial articles has increased in importance. Because of the controversial nature of some topics, the narrative in a Wikipedia article may contain misleading information that stops a neutral consensus emerging. Prior work in this area has established insights such as the predictability of controversy from editor behaviour \cite{wiki3}, such as deletions, reversions, and statistics from the collaboration network, prediction of article quality taking insights from multiple models \cite{Wu2012},
and interactions between users, bots, admin and pages \cite{wiki4}. There has also been a number of different types of network developed to assess Wikipedia articles, including collaboration networks \cite{brandes2009network} that capture the positive or negative relationship between editors, edit networks that capture "undoing'' of edits by a third party \cite{kittur2007he} and affiliation networks \cite{kane2016research}.

\begin{figure}[!t]
\centering
 \begin{tikzpicture}[
            > = stealth, 
            shorten > = 1pt, 
            auto,
            node distance = 3cm, 
            semithick 
        ]

        \tikzstyle{every state}=[
        	circle,
            draw = black,
            thick,
            fill = white,
            line width = 1mm,
            minimum size = 4mm
        ]

        \node[state, line width=0.2mm] (1) {Editor A};
        \node[state, line width=0.2mm] (2) [below left of=1] {Editor B};
        \node[state, line width=0.2mm] (3) [below right of=1] {Editor C};
        \node[state, line width=0.2mm] (4) [below left of=3] {Editor D};

        \path[->] (1) edge node {} (2);
        \path[->] (2) edge node {} (4);
        \path[->] (4) edge node {} (1);
        \path[->] (1) edge node {} (3);
        \path[->] (3) edge node {} (1);

    \end{tikzpicture}
    \caption{A network generated using the editor order $A, B, D, A, C, A$, from newest edit to oldest.  Each editor is characterised by a letter $A$ to $D$, and each occurrence in the list marks a single revision of the article by the corresponding author. An edge is formed between the current editor and its adjacent neighbour in the sequence, forming the directed edges $(A, B), (B, D), (D, A), (A, C), (C, A)$.
    }
    \label{fig:simple1}
  \end{figure}
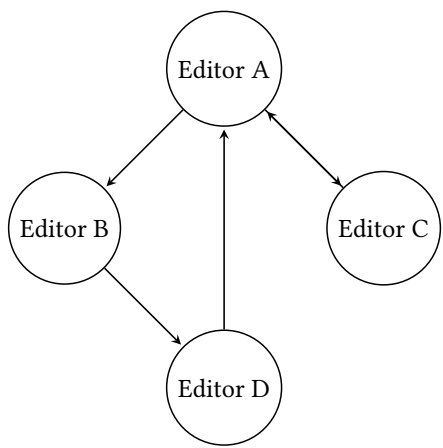

Our focus in this paper is to further understand the relationship between small groups of editors, as induced by their editing sequences, by using a {\em revision network}. This does not require information on the nature of the editing undertaken - it simply captures the ordering in which editing occurs and is therefore a simple metric to infer. Editors are represented by nodes and a directed edge from node $A$ to $B$ indicates that "\emph{Editor A edits the article after Editor B}'' (see Figure \ref{fig:simple1}). From this, we seek to determine the extent to which controversial articles exhibit a distinctive signature relative to those that are deemed non-controversial. 

There has already been some consideration of revision networks in the literature \cite{iba2010analyzing,keegan2012staying}, where more recently the emphasis \cite{wu2015integration} has been to combine them with other network representations. However, given the fundamental nature of revision networks, it is interesting to question the extent to which they hold sufficient information to characterise controversial Wikipedia articles. Currently this is not well-understood, and motivates our work.

\begin{figure*}[!t]
\centering
\includegraphics[width=\textwidth]{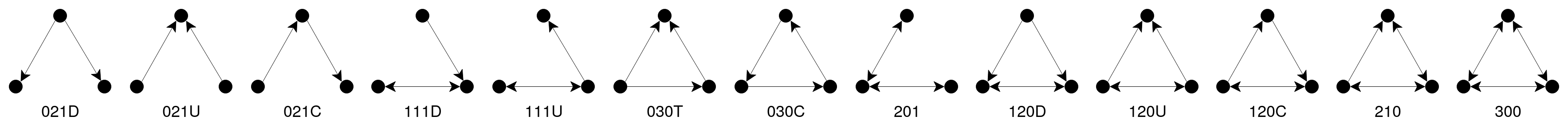}
\caption{13 possible combinations of connected triads in directed networks.}
\label{fig:triads}
\end{figure*}

\subsection{Hypotheses}
We hypothesise that {\em interaction differences between small sub-groups of Wikipedia editors is sufficient to distinguish between controversial articles and non-controversial articles}. To address this we consider the extent to which revision networks of Wikipedia articles have different local induced substructures based on their having controversial classifications. Our approach is based on techniques from complex networks \cite{milo2002network,milo2004superfamilies}, that have been successful in classifying diverse and complex biological networks based on their latent induced subgraphs. 

To achieve this at scale, and in contrast to previous literature \cite{wiki2, wiki3, wiki4}, we assess a relatively large sample of Wikipedia articles, involving over 21,000 Wikipedia articles, by determining their subgraph ratio profiles. Each such profile represents the under and over representation of induced triads in the revision network of a Wikipedia article using 13 dimensions of connected triads, while also normalising for differences in network size. 

Understanding the relationship between Wikipedia articles is important, as it allows us to determine the extent to which they may cluster. The sample we are considering represents 21,631 articles each represented in 13 dimensions.  Therefore we perform dimensionality reduction, and project the subgraph ratio profiles into lower dimensional spaces. This allows us to examine the relationship both within and between controversial and non-controversial articles. 

The results draw attention to distinctive clustering patterns concerning controversial Wikipedia articles. 
We further assess the results by examining the correlation with a range of variables, allowing to understand the role of substructures. The findings reaffirm that the sequence of editing provides an important mechanism to understand Wikipedia articles, independent of an article's topic and without recourse to semantic analysis. 

\section{Related Work}
Understanding the content of crowd-sourced platforms such as Wikipedia and the behaviour of their contributors is of wide research interest \cite{Surowiecki2005}. 
Wikipedia represents a dynamic network of articles with a structure resembling that of the World Wide Web \cite{wiki_web}, whereby dominant articles act as connectivity hubs. Dynamics also exist within the formation and maintenance of individual Wikipedia articles, through open and collaborative editing. 

Interactions between editors range from positive to negative, where debates and arguments lead to different patterns of revision (e.g., \cite{wiki3, laniado2011wikipedians}), capturing behaviours such as vandalism \cite{potthast2008automatic} and the propagation of disinformation \cite{kumar2016disinformation}. Characterising articles and contributors through revisions provides a means for Wikipedia to manage and review its content. This is potentially labour intensive and has led to interest in creating and exploiting methods to detect issues (e.g.,\cite{adler2011wikipedia,sepehri2015identifying}).

Controversial articles have become an increasing point of focus, and characterised as such by Wikipedia. Automated methods for classifying articles have received much attention (e.g., \cite{rad2012identifying, Wu2012, wu2015integration}. The associated revision log for Wikipedia articles has been shown to provide a basis to examine potential controversy through examining the collaborative behaviour of individual editors within an article \cite{sepehri2015identifying} or across multiple articles \cite{wu2011characterizing}.
An article's revision log identifies the structure underlying temporal interactions \cite{wu2015integration}, and provides insight into how articles and contributors' habits may evolve over time \cite{wiki4}.
Features from the aggregation of this, such as number of edits, revision, and previous version restorations have been shown to correlate (e.g., \cite{sepehri2015identifying}).

Treating the revision log as a network between editors \cite{wiki2} has been shown to provide additional useful features using graph theory and social network analysis techniques (e.g., \cite{rad2012identifying}). This has ranged from global features such as the degree distribution (e.g., \cite{sepehri2015identifying}), through to analysis of local sub-structures concerning the articles with which editors interact (e.g., \cite{wu2011characterizing}).

However, there has been little investigation of controversial articles based on the under or over representation of local-substructures. Called network motif analysis, this approach originated from biology \cite{milo2004superfamilies,braines2018role}, and has been used to good effect in characterising other complex networks, including technology (e.g., \cite{OCallaghan2012}). In terms of Wikipedia, motif analysis has been used to determine how articles point to each other \cite{wiki_web} and in assessing interactions between editors and different Wikipedia articles \cite{wu2011characterizing}. Our contribution is two-fold: firstly to use motif analysis to understand the fundamental revision networks, and secondly to consider the role of motifs in differentiating between controversial and non-controversial articles.

\section{Methods}



\subsection{Dataset}

We collected revision history logs and article meta-data of a sample of Wikipedia articles ($N=21,631$) through Wikipedia's web API\footnote{\url{https://en.wikipedia.org/w/api.php}}. The revision logs contain time-series events and meta-data attributing the revision to a particular user at a given time. 
Within this set of articles, a subset ($N=2,661$) are considered to be 'controversial' as they are listed in Wikipedia's "List of controversial issues''$^1$. The remaining articles ($N=18,970$) are random articles that are not contained in the controversial issue list to serve as a basis for comparison. These were taken from an original sample of 23,000 articles (20,000 non-controversial and 3,000 controversial), from which articles were removed if they did not contain sufficient information for motif analysis. 

\begin{figure*}[!t]

\centering
\subfloat[A revision network taken from the non-controversial \textit{The Web Conference} Wikipedia article.]{
    \includegraphics[width=0.35\textwidth]{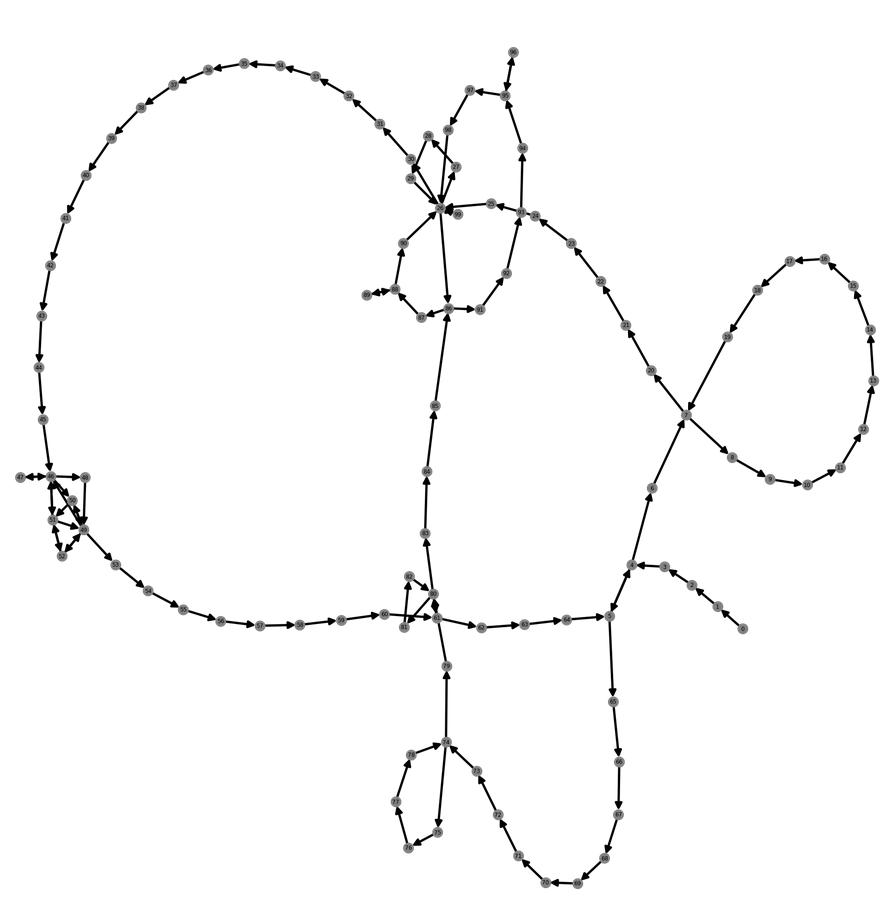}
    \label{fig:wiki1}
}
\qquad
\qquad
\qquad
\qquad
\subfloat[A revision network taken from the controversial \textit{Brexit} Wikipedia article. The large volume of editors produces a highly dense cluster of interactions.]{
    \includegraphics[width=0.35\textwidth]{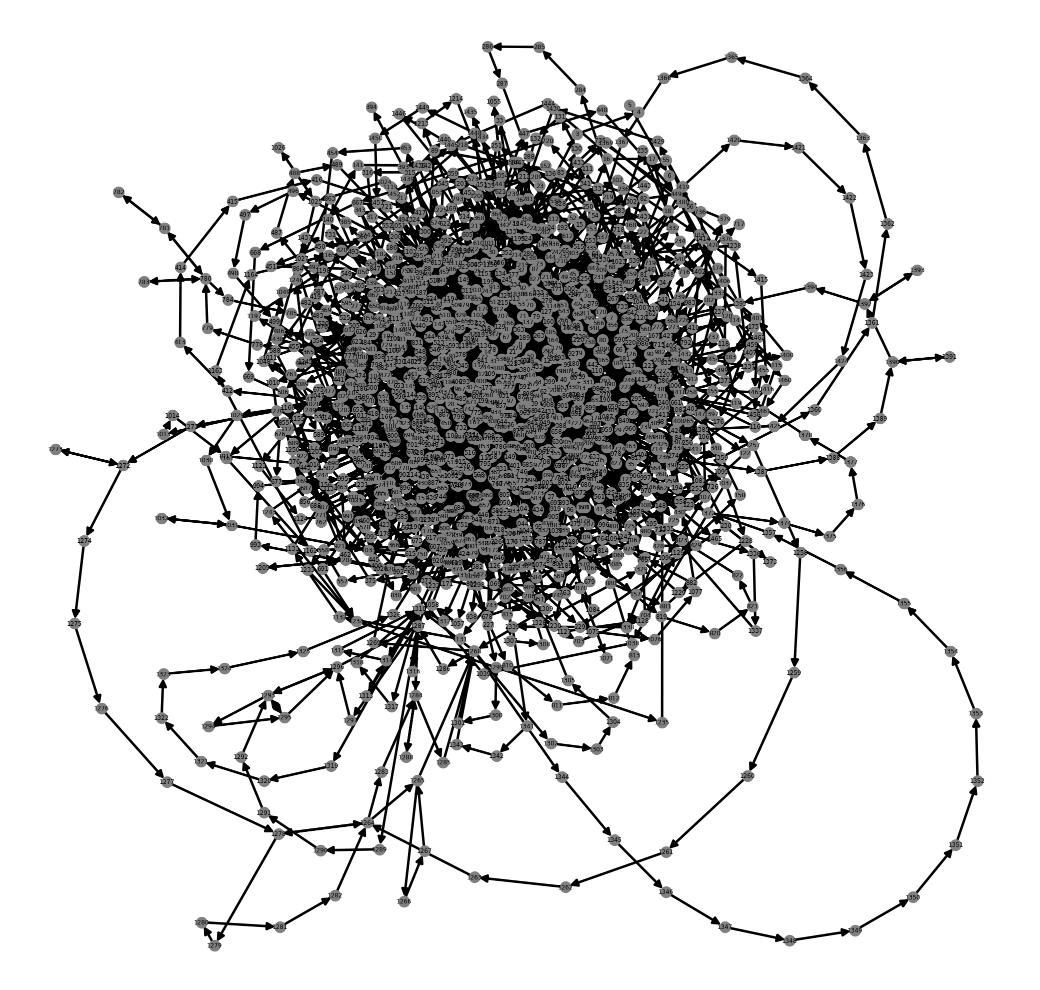}
    \label{fig:wiki2}
}

\caption{Revision network of two articles - one non controversial and one controversial.}
\label{fig:rn}
\end{figure*}

\subsection{Network Construction}

For each article, we construct a revision network (in the same manner as \cite{keegan2012staying, wu2015integration}) where nodes represent unique editors and directed edges show that an editor added a revision after another editor. We traverse the revision log list to build a network that spans the article's lifetime, adding nodes and edges as they appear in each event. Specifically let the revision network of a Wikipedia article be defined by $G=(V,E)$, where each editor is represented by a node $v \in V$. An edge $(v_i,v_j) \in V$ indicates that editor $v_i$ edits the article after editor $v_j$. This excludes self-loops and editor $v_i$ editing after editor $v_j$ multiple times does not result in a multi-edge. 

A simple example can be found in Figure \ref{fig:simple1}, which describes how the network is constructed. 
Two extreme examples from the dataset shown in Figures \ref{fig:wiki1} and \ref{fig:wiki2}. This form of representation is potentially useful as large articles do not typically follow a linear or incremental structure. For example, it is highly likely that users will restore work back to an earlier revision should a revision become vandalised or irrelevant. Editors are likely to refer back to previous editor's work. These behaviours result in complex sequential patterns that are captured through revision networks.




\subsection{Network Motif Analysis}

Network motif analysis focuses on determining the under or over representation of induced subgraphs \cite{milo2002network,braines2018role}, as compared to an alternative sample of graphs (i.e., a null-model that acts a relevant baseline for comparison). Dyads, triads and tetrads are often considered as induced subgraphs such over and under-represented induced subgraphs are called network {\em motifs} or {\em anti-motifs}. 
In this analysis, we examine each article's revision network using triads, representing how all possible triples of editors may sequentially interact. Triads are sufficiently large enough to capture both direct and indirect reciprocity between editors, while not being of a scale that is impeded by combinatorial complexity - there are 13 possible connected triads, shown in Figure \ref{fig:triads}. The coded names listed in Figure \ref{fig:triads} are provided as part of the convention used in the triad census algorithm \cite{BATAGELJ2001237}.

For each article, we calculate the subgraph ratio profile (SRP) as defined by Milo et al \cite{milo2004superfamilies}. This accounts for variations in network size. This is achieved by determining the relative abundance of each type of triad compared to random graphs generated by the null-model. For each type of triad $i$, we firstly calculate $\Delta_i$: 

\begin{equation}
    \Delta_i =  \frac{(Nreal_i - <Nrand_i>)}{(Nreal_i + <Nrand_i>) + \epsilon}
\end{equation}
where $Nreal_i$ is the number of such triads observed in the graph under observation, and $Nrand_i$ is the average frequency of such triads seen across the sample of networks composed by the null model. In this case the null model uses 100 random graphs with the same number of nodes and edges as the graph under observation. The value of $\epsilon$ is set to $4$ \cite{milo2004superfamilies,tu2018network} to prevent the result from being misleadingly large when a subgraph rarely appears. This process is repeated for each triad $i$ and normalised across triads to form the {\em subgraph ratio profile} (SRP) for a given network. The $i^{th}$ SRP, denoted $SRP_i$, denotes the extent of under or over representation of the triad $i$, and is defined by:
\begin{equation}
    SRP_i = \Delta_i / (\sum \Delta_i^2)^{1/2}.
\end{equation}

\subsection{Article comparison and dimensionality reduction}

The SRP composed for each article provides a 13-dimensional vector whose components indicate the extent of triad representation relative to networks in the null model. To assess these collectively, we use principal component analysis (PCA) that allows the SRPs to be considered in a lower dimensional space. We apply three and two dimensional PCA. 
Finally, we make comparison with a number of external variables (number of editors/nodes, age of article, and revision rate) to understand potential correlations with motifs.

\section{Results}

\subsection{Subgraph Ratio Profiles}

\begin{figure}[!t]
\centering
\includegraphics[width=0.45\textwidth]{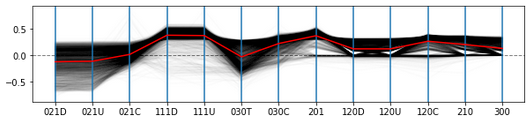}
\caption{Subgraph ratio profiles of all controversial articles. Average displayed in red.}
\label{fig:sp1}
\end{figure}

\begin{figure}[!t]
\centering
\includegraphics[width=0.45\textwidth]{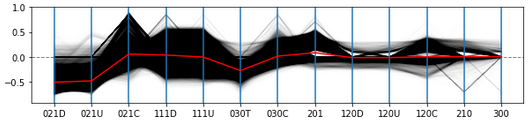}
\caption{Subgraph ratio profiles of all non-controversial articles. Average displayed in red.}
\label{fig:sp3}
\end{figure}

We first examine the SRPs that arise from both controversial and non-controversial articles in isolation (Figures \ref{fig:sp1} and \ref{fig:sp3} respectively). 
To find significant triads, we use a cutoff value of +0.3 and -0.3. We determine that controversial articles are strongly represented by triads \verb|111D|, \verb|111U| and \verb|201|, which attain average {\em SRP} scores of 0.382, 0.375, 0.372, with relatively low dispersion (SDs of 0.136, 0.149 and 0.124 respectively). Interestingly, together these represent a chain of three nodes, where one edge is reciprocated, with the other edge covering all possible directional types (i.e., reciprocated, directed in, directed out).



In contrast, the results for the non-controversial articles provide a different profile. Here \verb|021D| and \verb|021U| are significantly under represented (average {\em SRP} scores of -0.511, -0.485), albeit with higher standard deviations present (SDs of 0.192, 0.2). Interestingly these anti-motifs (\verb|021D| and \verb|021U|) relate to a lack of subgraphs where directed edges either emanate from or are received by a single node in the triad. Such configurations relate to the role of a mediating editor that may be presented with or respond to the editing of others. In other words, such mediators have a reduced role in non-controversial articles.  


\begin{figure}[!t]
\centering
\includegraphics[width=0.45\textwidth]{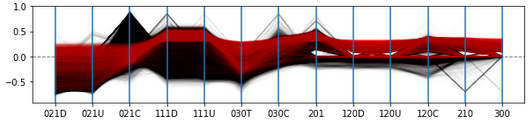}
\caption{Subgraph ratio profiles of all articles, with controversial articles displayed in red and non-controversial in black.}
\label{fig:sp2}
\end{figure}

The comparison between these subgraph ratio profiles is shown in Figure \ref{fig:sp2}. These profiles are quite distinct. We also calculate the Pearson correlation coefficient for each distinct pair of articles in three groups - controversial articles, non-controversial articles and all articles. 
Controversial articles provide the greatest correlation to each other (M=$0.41193$, SD=$0.41262$). Non-controversial articles have a lower mean correlation (M=$0.37498$, SD=$0.36749$) which is similar to the result when considering all articles together (M=$0.37569$, SD=$0.36844$).

\subsection{Principal Component Analysis}

\begin{figure}[!t]
\centering
\includegraphics[width=0.425\textwidth]{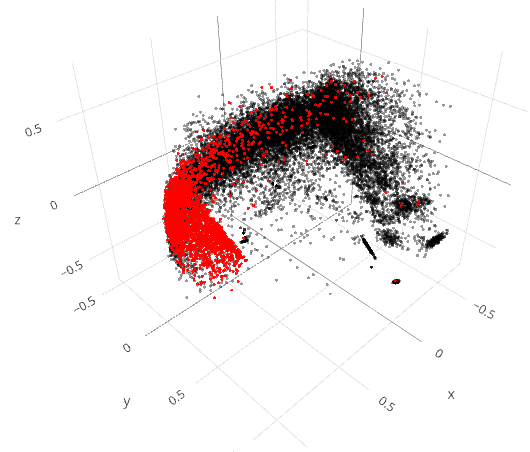}
\caption{PCA scatter plot repression of the 13-point feature vector in 3D clustering space.}
\label{fig:pca1}
\end{figure}

We analyse the 21,631 subgraph ratio profiles to determine the relationship in terms of relative clustering. We apply PCA in order to reduce the 13 dimensions of the SRPs down to a more manageable form. We initially project the SRPs into 3-dimensional space for clarity, as seen in Figure \ref{fig:pca1}. This presents a distinctive region where controversial articles are dominant. This provides evidence for a distinction between the controversial and non-controversial articles, consistent with the variation in motifs identified in the previous section.

The PCA coefficients (Table \ref{coef}) that define the three dimensions reveals that principal component one ($x$ axis in Figure \ref{fig:pca1}) primarily depends on triads \verb|111D| and \verb|111U|. Principal component two ($y$ axis in Figure \ref{fig:pca1}) primarily depends on triad \verb|021C|. The third principal component primarily depends on triads \verb|021C| and \verb|030C|. However, we also note that when represented in the three dimensional space, the revision networks have limited dispersion in the third dimension (i.e., vertical dimension as plotted). 

\begin{table*}[!t]
\caption{PCA coefficients displaying the strongest triads}
\label{coef}
\begin{tabular}{c|c|c|c|c|c|c|c|c|c|c|c|c|c|}
\cline{2-14}
                                    & \textbf{021D} & \textbf{021U} & \textbf{021C} & \textbf{111D} & \textbf{111U} & \textbf{030T} & \textbf{030C} & \textbf{201} & \textbf{120D} & \textbf{120U} & \textbf{120C} & \textbf{210} & \textbf{300} \\ \hline
\multicolumn{1}{|c|}{\textbf{PC-1}} & 0.332         & 0.321         & -0.126        & 0.509         & 0.522         & 0.145         & 0.205         & 0.318        & 0.078         & 0.079         & 0.204         & 0.117        & 0.065        \\ \hline
\multicolumn{1}{|c|}{\textbf{PC-2}} & 0.427         & 0.438         & 0.643         & -0.278        & -0.242        & 0.207         & 0.104         & -0.052       & 0.059         & 0.06          & 0.047         & 0.074        & 0.06         \\ \hline
\multicolumn{1}{|l|}{\textbf{PC-3}} & 0.074         & 0.127         & -0.609        & -0.398        & -0.332        & 0.136         & 0.441         & 0.115        & 0.131         & 0.130         & 0.19          & 0.167        & 0.112        \\ \hline
\end{tabular}
\end{table*}

\begin{figure}[!t]
\centering
\includegraphics[width=0.425\textwidth]{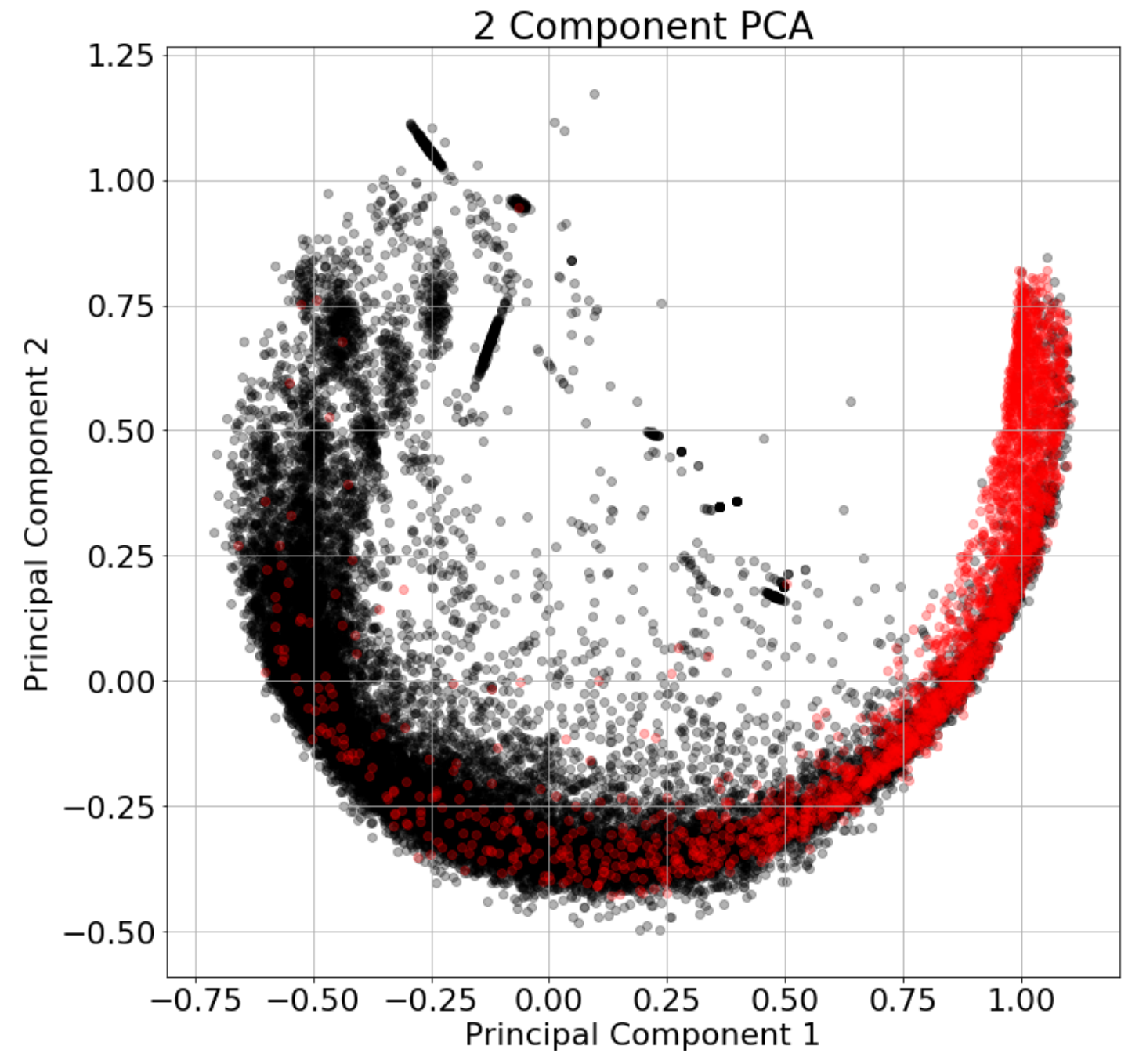}
\caption{PCA scatter plot repression of the 13-point feature vectors in 2D clustering space.}
\label{fig:pca_2}
\end{figure}

Calculating the percentage of explained variance by principal component confirms that the first principal component produces 53.7\% of the shared variance, the second produces 22.3\% and the third produces the least with 6.7\%. This confirms that the third principal component provides a limited contribution to representation of the total variance across the significant ratio profiles. This supports representation through two principal components, as plotted in Figure \ref{fig:pca_2}, with the relative composition of each principal component being near identical to PC-1 and PC-2 in Table \ref{coef}. As anticipated, this is similar dependency on the first and second primary components in the three dimensional representation. 

Representation in two dimensions further clarifies the distinction between controversial and non-controversial revision networks. In particular, from Figure \ref{fig:pca_2} we note that both classes of article exhibit a similar maximum and minimum range against principal component two, which is primarily defined by the linear path between three nodes (\verb|021C|). However, it is the variation in the first principal component, dominated by \verb|111D| and \verb|111U|, which represent linear paths with reciprocation on one edge, that differentiate the non-controversial from controversial. High values in principal component one correlate with controversial articles - in other words,  controversial articles exhibit more reciprocation on top of possible linear paths.


\subsection{Additional Metrics}
\begin{figure}[!t]
\centering
\includegraphics[width=0.45\textwidth]{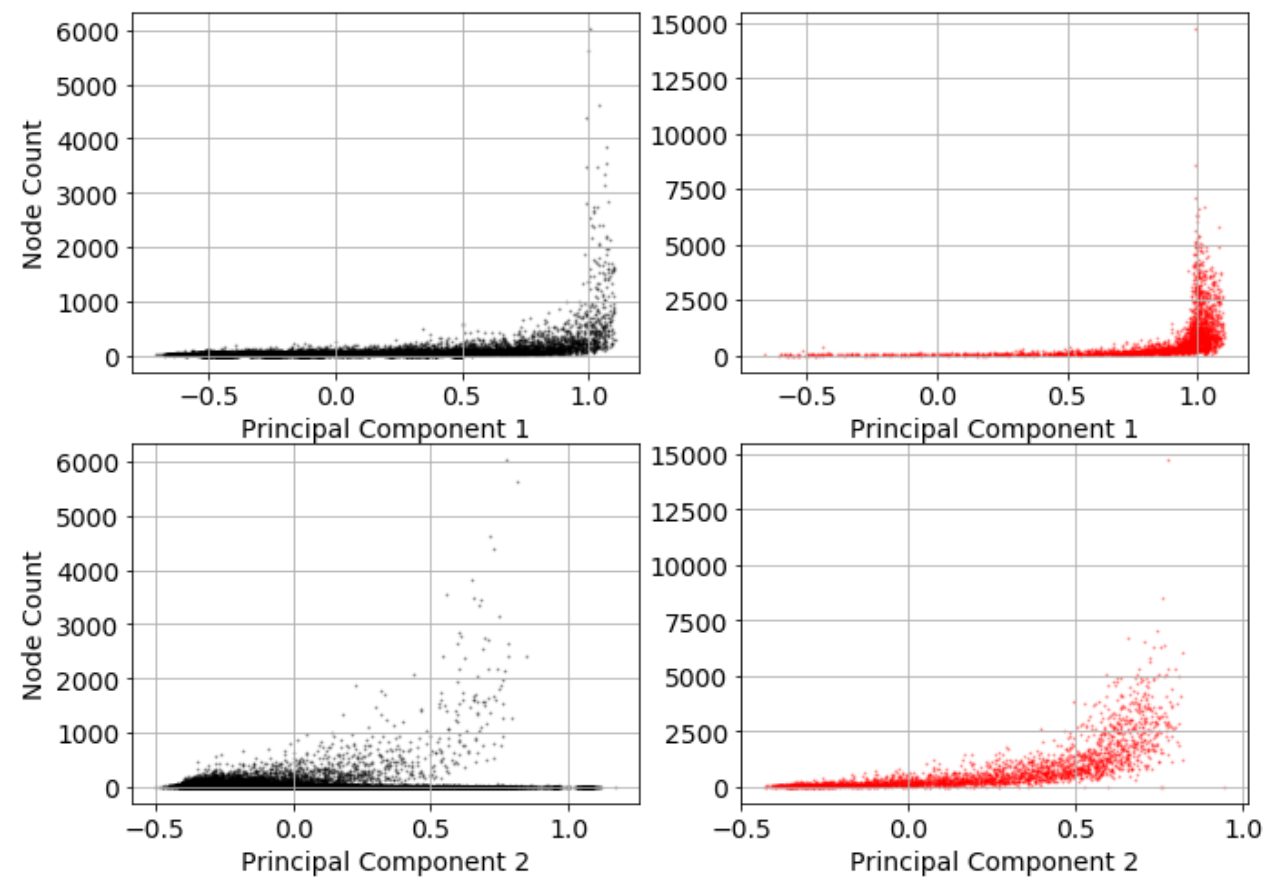}
\caption{Scatter plot of each principal component combined with node count for both non-controversial and controversial articles.}
\label{fig:pc_nc}
\end{figure}

\begin{figure}[!t]
\centering
\includegraphics[width=0.45\textwidth]{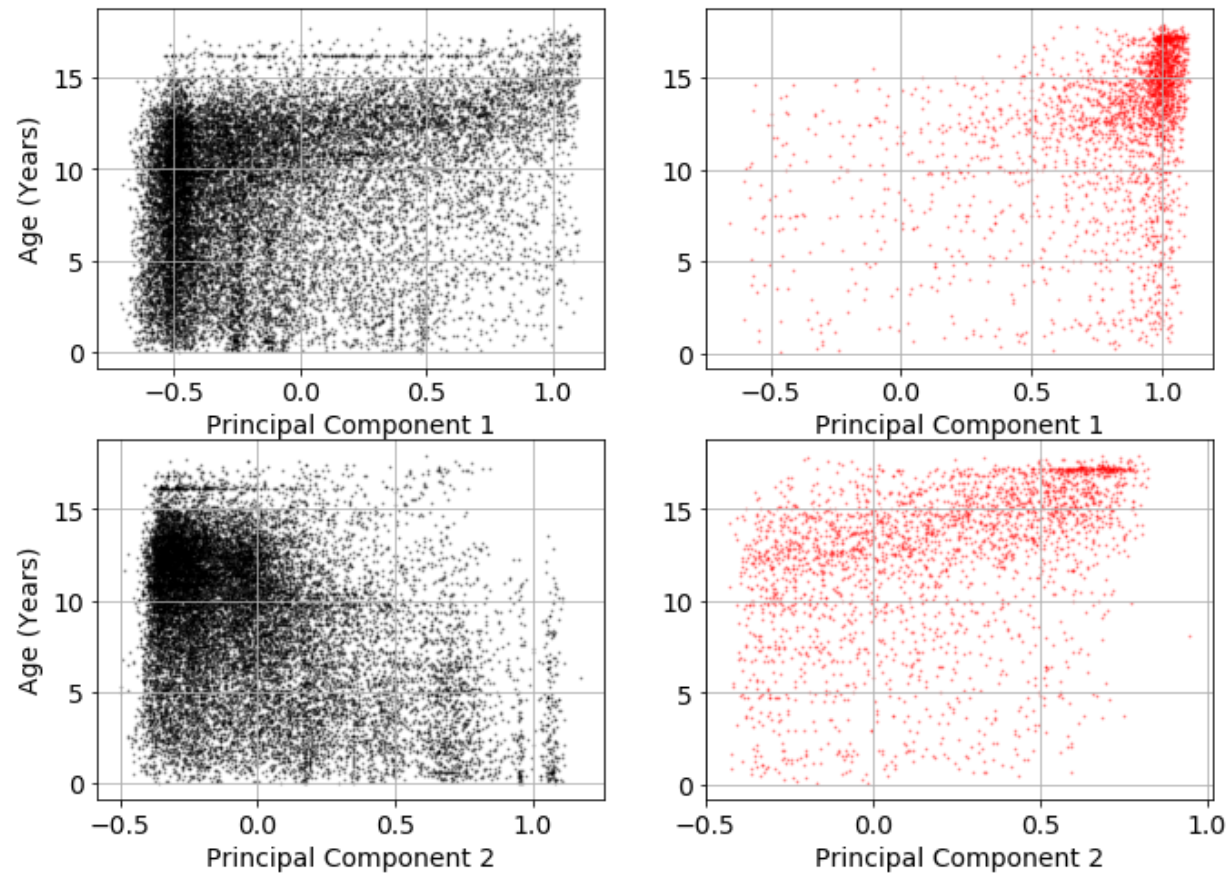}
\caption{Scatter plot of each principal component combined with article age for both non-controversial and controversial articles.}
\label{fig:pc_age}
\end{figure}

\begin{figure}[!t]
\centering
\includegraphics[width=0.45\textwidth]{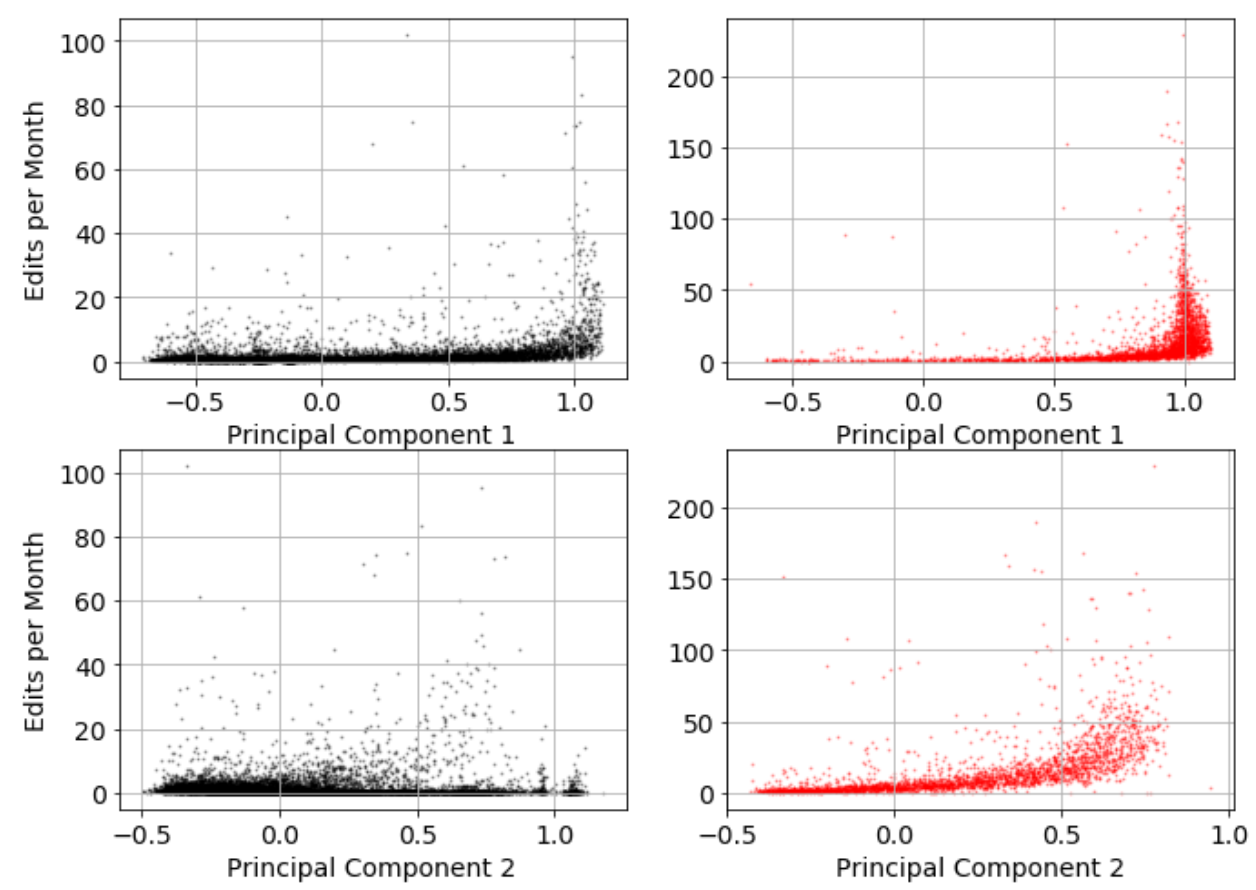}
\caption{Scatter plot of each principal component combined with edit rate (mean number of edits per month) for both non-controversial and controversial articles.}
\label{fig:pc_er}
\end{figure}

We examine the relationship between revision networks and primary external variables (number of editors/nodes, age of article, and revision rate) using motifs. Specifically, using the dimensions of two-dimensional PCA analysis, we examine the correlation with the external variables, and how this differs between controversial and non-controversial articles. The results are shown in Figures \ref{fig:pc_nc}, \ref{fig:pc_age} and \ref{fig:pc_er}. 

The greatest differences between controversial and non-controversial articles occur with respect to article age (Figure \ref{fig:pc_age}). Here, controversial articles with high age cluster against high values of principal component one, and to some extent this occurs for principal component two. This contrasts against the clustering seen for non-controversial articles. 

Figure \ref{fig:pc_age} also shows that controversial articles have a tendency to be older. The high values in principal component one and two which align with dense clustering of controversial articles show that while such articles accumulate the linear revision path between authors (\verb|021C| which dominates principal component two), controversial articles also accumulate instances of linear paths where one edge is reciprocated (i.e., \verb|111D| and \verb|111U| which dominate principal component one). 

\section{Discussion}
Motif analysis of revision networks gives insight into how the temporal editing relationship between small groups of Wikipedia authors create signatures that allow controversial articles to be distinguished. In contrast to previous work, we have investigated this using a relatively large sample of Wikipedia articles, where distinct patterns emerge. This provides strong support for our hypothesis, and reaffirms the importance of the revision network as a simple but fundamental element in editing Wikipedia.

Through motif analysis, we identify that reciprocation on linear paths among triads of editors in the revision network is over represented in controversial Wikipedia articles. These motifs are defined by the triads \verb|111D|, \verb|111U| and \verb|201|. In contrast, the revision networks from non-controversial articles exhibit two anti-motifs, involving the under representation of triads involving two directed edges either arriving at or emanating from a mediating node (triads \verb|021D| and \verb|021U|). These motifs and the underlying subgraph ratio profiles represent an unusual and distinctive profile that we believe represent distinctive "super-families'' beyond those seen in other technologically related networks, such as the world wide web \cite{wiki_web}. 

Performing dimensionality reduction upon the subgraph ratio profiles from each revision network allows us to further understand the relationships between Wikipedia articles. Our analysis shows that the structure of the data is amenable to reduction to two dimensions, where the principal components are dominated by triads \verb|111D| and \verb|111U|  in the first component, and mainly \verb|021C| in the second component, but with lesser contributions from triads \verb|021D| and \verb|021U|. 

The results from two-dimensional principal analysis are illuminating - the dominant triads in both components, as listed above, correspond to linear paths, i.e., open triads which represent sequences of editing without indirect reciprocity. The extent and format of reciprocated (i.e., bidirectional) edges on these open triads is sufficient to define the two principal components. The dominant triads in the first principal component each involve reciprocation on one edge, where as interestingly, in the second principal component, the dominant triads are open triads with no reciprocated edges. From this we deduce that short paths, rather than short loops of editing that represent indirect reciprocity, are important features in characterising Wikipedia revision networks. 

We also observe through two-dimensional principal component analysis that it is the first principal component that strongly distinguishes between the revision networks of controversial and non-controversial articles. The dominant triads defining this capture the extent of direct reciprocation being present. Finally, through consideration of the principal components against additional external variables, we find in particular that article age plays a role in distinguishing the controversial articles. High values in both principal components aligns with strongest clustering of controversial articles, which is not the case for non-controversial articles. 

\section{Conclusion}
The analysis has given insights into the structure underlying revision networks from Wikipedia articles, and has shown that a relatively small number of features, in terms of substructures in revision networks, characterise controversial Wikipedia articles. The results have identified key clusters of editorial interactions to this effect, in support of the hypothesis. These are distinctive and indicate that the revision networks for controversial and non-controversial Wikipedia articles have differentiated subgraph ratio profiles. Our study gives understanding as to how prediction or classification of articles can be enhanced using the latent structures relating to editor behaviour. This also reaffirms the importance of the revision network as a simple but useful representation for assessment of Wikipedia articles.




\section*{Acknowledgements}

\small{
This research was sponsored by the U.S. Army Research Laboratory and the U.K. Ministry of Defence under Agreement Number W911NF-16-3-0001. The views and conclusions contained in this document are those of the authors and should not be interpreted as representing the official policies, either expressed or implied, of the U.S. Army Research Laboratory, the U.S. Government, the U.K. Ministry of Defence or the U.K. Government. The U.S. and U.K. Governments are authorized to reproduce and distribute reprints for Government purposes notwithstanding any copyright notation hereon.
}

\balance{}

\bibliographystyle{ACM-Reference-Format}
\bibliography{main}

\balance{}


%

\end{document}